\def\A{\leavevmode\setbox0\hbox{A}\lower1.4ex\hbox to\wd0
{\hss`}\kern-.9\wd0A}
\def\E{\leavevmode\setbox0\hbox{E}\lower1.4ex\hbox to\wd0
{\hss`\/}\kern-.9\wd0E}
\def\a{\leavevmode\setbox0\hbox{a}\lower1.4ex\hbox to\wd0
{\hss`\/}\kern-\wd0a}
\def\e{\leavevmode\setbox0\hbox{e}\lower1.4ex\hbox to\wd0
{\hss`\/}\kern-\wd0e}

\font\symb=msam7
\def\znakr{\raise1.5pt\hbox{\symb\char66\kern-2pt\char74}}
\def\znakl{\raise1.5pt\hbox{\symb\char73\kern-2pt\char67}}

\def\normalsize{
\setlength{\textheight}{23cm}
\setlength{\textwidth}{15cm}
\setlength{\topmargin}{-2.0cm}
\setlength{\hoffset}{-0.5cm}
\setlength{\leftmargin}{-1cm}
\setlength{\rightmargin}{2.0cm}}

\documentstyle[12pt]{article}
\normalsize

\begin{document}
\title{On Propagator of Notoph}
\author{M. Bakalarska, W. Tybor\thanks{Supported by
\L{}\'od\'z University Grant No. 505/581} \\
Department of Theoretical Physics \\
University of \L \'od\'z \\
ul. Pomorska 149/153, 90-236 \L \'od\'z, Poland}
\date{}
\maketitle
\setcounter{section}{0}
\setcounter{page}{1}
\begin{abstract}
It is shown that the notoph propagator in the noncovariant \\
longitudinal gauge is equivalent to the covariant Feynmann - like \\
propagator.
\end{abstract}

\newpage

{\bf 1.} Let us consider the interaction of the notoph \cite{1}
( i.e. the scalar particle described by the antisymmetric tensor field 
$B^{\mu\nu} = - B^{\nu\mu}$) with the external current $j^{\mu\nu}
= - j^{\nu\mu}$. The Lagrangian is 
\begin{equation}
  {\cal L} = - \frac{1}{2} G_{\mu} G^{\mu} + \frac{1}{2} j_{\mu\nu}
  B^{\mu\nu},
  \label{1}
\end{equation}
where $G^{\mu} = \partial_{\nu} B^{\nu\mu}$. The theory is invariant 
under the gauge transformation 
\begin{equation}
  \delta B^{\mu\nu} = \varepsilon^{\mu\nu\alpha\beta} \partial_{\alpha}
  \lambda_{\beta},
  \label{2}
\end{equation}
(where $\lambda^{\beta}$ is an arbitrary 4 - vector function), if
the current obeys the conservation law 
\begin{equation}
  \varepsilon^{\mu\nu\alpha\beta} \partial_{\nu} j_{\alpha\beta} = 0.
  \label{3}
\end{equation}
The vector $G^{\mu}$ is invariant under the transformation (\ref{2})
and it is a strength vector in the notoph theory.

The free action can be rewritten in the form 
\begin{equation}
  \int d^{4} x (- \frac{1}{2} G_{\mu}G^{\mu}) = \frac{1}{4} \int
  d^{4} x B_{\mu\nu} L^{\mu\nu\alpha\beta} B_{\alpha\beta},
\end{equation}
where the Lagrangian operator $L^{\mu\nu\alpha\beta}$ is
\begin{equation}
  L^{\mu\nu\alpha\beta} = \Box \Pi^{\mu\nu\alpha\beta}
\end{equation}
and
\begin{equation}
  \Pi^{\mu\nu\alpha\beta} = \frac{1}{2 \Box} \left(g^{\mu\alpha}
  \partial^{\nu} \partial^{\beta} + g^{\nu\beta} \partial^{\mu} 
  \partial^{\alpha} - g^{\mu\beta} \partial^{\nu} \partial^{\alpha}
  - g^{\nu\alpha} \partial^{\mu} \partial^{\beta}\right).
\end{equation}
The operator $\Pi^{\mu\nu\alpha\beta}$ has the symmetry properties 
of a Riemann tensor.\\
There is no inverse operator to the Lagrangian one because $\Pi^{\mu
\nu\alpha\beta}$ is a projector operator
\[
  \Pi^{\mu\nu\alpha\beta} {\Pi_{\alpha\beta}}^{\sigma\lambda} = 
  \Pi^{\mu\nu\sigma\lambda}.
\]
The origin of the Lagrangian singularity is the gauge freedom of the 
theory. To fix the gauge we impose the gauge condition 
\begin{equation}
  \varepsilon^{\mu\nu\alpha\beta} \partial_{\nu} B_{\alpha\beta} = 0.
  \label{7}
\end{equation}
We note that this condition does not contradict the field equation 
when we switch on the interaction with the external current.

Let us introduce the effective free action 
\begin{equation}
  \int d^{4} x \left[ - \frac{1}{2} G_{\mu} G^{\mu} + 
  \frac{1}{8 \alpha} (\varepsilon^{\mu\nu\sigma\lambda} \partial_{\nu}
  B_{\sigma\lambda})^{2} \right]  = \nonumber \\
  \frac{1}{4} \int d^{4} x B_{\mu\nu} L^{\mu\nu\sigma\lambda}_{(\alpha)}
  B_{\sigma\lambda},
\end{equation}
where the Lagrangian operator 
\begin{equation}
  L^{\mu\nu\sigma\lambda}_{(\alpha)} = (1 - \frac{1}{\alpha}) \Box 
  \Pi^{\mu\nu\sigma\lambda} + \frac{1}{\alpha} \Box E^{\mu\nu\sigma
  \lambda}
\end{equation}
can be inverted for any finite value of the parameter $\alpha$. The 
tensor
\begin{equation}
  E^{\mu\nu\alpha\beta} = \frac{1}{2} (g^{\mu\alpha} g^{\nu\beta} - 
  g^{\mu\beta} g^{\nu\alpha})
\end{equation}
is the unit tensor in a space of antisymmetric tensors.

The notoph propagator $D^{\mu\nu\alpha\beta}$ in the momentum space
is defined as the inverse matrix to the Lagrangian one 
\begin{equation}
  {L^{\mu\nu}}_{\alpha\beta} (k, \alpha) D^{\alpha\beta\sigma\lambda}
  = E^{\mu\nu\sigma\lambda}.
  \label{11}
\end{equation}
The general form of the propagator is 
\begin{equation}
  D^{\mu\nu\alpha\beta} = \frac{A}{k^{2}} E^{\mu\nu\alpha\beta} + 
  \frac{B}{k^{2}} \Pi^{\mu\nu\alpha\beta} (k).
\end{equation}
From Eq.\ (\ref{11}) we obtain 
\[
A = - \alpha,  \qquad  B = \alpha - 1.
\]
So, the notoph propagator is 
\begin{equation}
  D^{\mu\nu\alpha\beta} = - \frac{\alpha}{k^{2}} E^{\mu\nu\alpha
  \beta} + \frac{\alpha - 1}{k^{2}} \Pi^{\mu\nu\alpha\beta} (k).
  \label{13}
\end{equation} 
If $\alpha = 0$, we get the propagator in the gauge (\ref{7}). If 
$\alpha = 1$, we get the Feynmann - like propagator 
\begin{equation}
  D^{\mu\nu\alpha\beta} (k) = - \frac{1}{k^{2}} E^{\mu\nu\alpha\beta}.
  \label{14}
\end{equation}
We note that due to the current conservation, the effective propagator 
is the Feynmann - like one. Indeed, Eq.\ (\ref{3}) can be rewritten 
in the form 
\[
  \partial_{\nu} j_{\alpha\beta} + \partial_{\alpha} j_{\beta\nu} + 
  \partial_{\beta} j_{\nu\alpha} = 0
\]
and we get 
\begin{equation}
  \Box j^{\alpha\beta} = \partial^{\alpha} j^{\beta} -
  \partial^{\beta} j^{\alpha}
  \label{15}
\end{equation}
where$j^{\beta} \equiv \partial_{\nu} j^{\nu\beta}$. With the help of
Eqs \ (\ref{13}) and (\ref{15}) we obtain 
\[
  D^{\mu\nu\alpha\beta} j_{\alpha\beta} = - \frac{\alpha}{k^{2}}
  j^{\mu\nu} + \frac{\alpha - 1}{k^{2}} j^{\mu\nu} =
  - \frac{1}{k^{2}} j^{\mu\nu}.
\]
The result is identical if the propagator (\ref{14}) is used from 
the very beginning.

\bigskip {\bf 2.} In Ref. \cite{2} we have investigated the notoph theory in
the (noncovariant) longitudinal gauge. We recall the main result of 
this paper.

In some fixed Lorentz frame we introduce the 3 - dimension notation.
The decomposition of the notoph field is $B^{\mu\nu} = 
(\vec{E}, \vec{H})$, where
\[
B^{k0} \rightarrow E_{k}, \qquad  B^{kn} \rightarrow 
  \varepsilon_{knj} H_{j}, \qquad  k,n,j = 1,2,3.
\]
The gauge freedom is removed completely by imposing the longitudinal
gauge condition 
\[
rot \vec{E} = 0.
\]
In this gauge the propagators of the notoph are
\begin{eqnarray}
  D^{(E)}_{ij} (k) & = & \frac{k_{i} k_{j}}{{\mid\vec{k}\mid}^{2}}
  \frac{1}{k^{2}};\nonumber \\
  D^{(H)}_{ij} (k) & = & \left(\delta_{ij} - \frac{k_{i} k_{j}}
  {{\mid\vec{k}\mid}^{2}} \right) \frac{1}{{\mid\vec{k}\mid}^{2}}.
  \label{16}
\end{eqnarray}

In the Lorentz frame in that the gauge condition is imposed, we
introduce the 4 - vector $\eta^{\mu} = (1,0,0,0)$ \cite{3}.
Using this vector we can rewrite the propagators (\ref{16}) in the 
4 - dimension notation
\begin{eqnarray}
  D^{(E)}_{\mu\nu\alpha\beta} (k) & = & \frac{1}{k^{2}} {[(\eta k)^{2}
  - k^{2}]}^{-1} (\eta_{\mu} k_{\nu} - \eta_{\nu} k_{\mu}) 
  (\eta_{\alpha} k_{\beta} - \eta_{\beta} k_{\alpha});\nonumber \\
  D^{(H)}_{\mu\nu\alpha\beta} (k) & = & - {[(\eta k)^{2} - k^{2}]}^{-2}
  \left\{ g_{\mu\alpha} k_{\nu} k_{\beta} + g_{\nu\beta} k_{\mu}
  k_{\alpha} - g_{\mu\beta} k_{\nu} k_{\alpha} - g_{\nu\alpha}
  k_{\mu} k_{\beta} + \mbox{} \right. \nonumber \\
  && \mbox{} - (\eta_{\mu} \eta_{\alpha} k_{\nu} k_{\beta} + 
  \eta_{\nu} \eta_{\beta} k_{\mu} k_{\alpha} - \eta_{\mu} \eta_{\beta}
  k_{\nu} k_{\alpha} - \eta_{\nu} \eta_{\alpha} k_{\mu} k_{\beta}) + 
  \mbox{} \nonumber \\
  && \mbox{} + (\eta k)^{2} (g_{\mu\alpha} \eta_{\nu} \eta_{\beta} + 
  g_{\nu\beta} \eta_{\mu} \eta_{\alpha} - g_{\mu\beta} \eta_{\nu}
  \eta_{\alpha} - g_{\nu\alpha} \eta_{\mu} \eta_{\beta}) + \mbox{}
  \nonumber \\ 
  && \mbox{} - (\eta k) \left[g_{\mu\alpha} (k_{\nu} \eta_{\beta} + 
  k_{\beta} \eta_{\nu}) + g_{\nu\beta} (k_{\mu} \eta_{\alpha} + 
  k_{\alpha} \eta_{\mu}) + \right. \mbox{} \nonumber \\
  && \left. \left. \mbox{} - g_{\mu\beta} (k_{\nu} \eta_{\alpha} + k_{\alpha} 
  \eta_{\nu}) - g_{\nu\alpha} (k_{\mu} \eta_{\beta} + k_{\beta}
  \eta_{\mu}) \right] \right\}. 
\end{eqnarray}
Let us consider the exchange of the notoph between two currents
\begin{equation}
  A = \frac{1}{4} j^{\mu\nu} (-k) D_{\mu\nu\alpha\beta} (k, \eta)
  j^{\alpha\beta} (k)
  \label{18}
\end{equation}
where
\begin{equation}
  D_{\mu\nu\alpha\beta} (k, \eta) = D^{(E)}_{\mu\nu\alpha\beta} (k)
  + D^{(H)}_{\mu\nu\alpha\beta} (k).
\end{equation}
Using the current conservation law (\ref{3}) we can rewrite the 
amplitude (\ref{18}) with the help of the Feynmann - like propagator
(\ref{14})
\begin{equation}
  A = \frac{1}{2} j^{\mu\nu} (-k) D_{\mu\nu\alpha\beta} (k) j^{\alpha
  \beta} (k).
  \label{20}
\end{equation}
So, the notoph propagator in the longitudinal gauge is equivalent 
to the covariant Feynmann - like propagator.

\bigskip {\bf 3.} Let us note that the covariant form (\ref{14}) of the notoph
propagator can be confirmed in another way. Performing the canonical 
analysis of the theory described by the Lagrangian (\ref{1}), we get 
the physical Hamiltonian of the interaction
\[
  {\cal H}_{int} = - j_{2Lk} B^{0k}_{L} + \frac{1}{2} j^{i}_{1T}
  \frac{1}{\Delta} j_{1Ti},
\]
or, in the momentum space
\begin{equation}
  {\cal H}_{int} = - j_{2Lk} (-k) B^{0k}_{L} (k) - \frac{1}{2}
  {\mid\vec{k}\mid}^{-2} j^{i}_{1T} (-k) j_{1Ti} (k)
\end{equation}
where
\[
  j^{i}_{1} \equiv j^{0i}, \qquad j^{i}_{2} = - \frac{1}{2} 
  \varepsilon^{imn} j_{mn},
\]
and the mark $L(T)$ denotes the longitudinal (transversal) part of 
the 3 - vector. Using the standard methods of the S - matrix 
formalism  \cite{3}, with the help of this Hamiltonian, we can
calculate  the amplitude of the exchange of the gauge field between
two external currents. We get 
\[
  A = - \frac{j_{2Li} (-k) j^{i}_{2L} (k)}{k^{2}} - \frac{j_{1Ti} 
  (-k) j^{i}_{1T} (k)}{{\mid\vec{k}\mid}^{2}}.
\]
It is the amplitude (\ref{20}) rewritten in the component form with 
the help of the current conservation law (\ref{3}).

{\bf 4.} Let us consider in es
 detail the case of the strong
conserving current \cite{4}
\begin{equation}
  j^{\mu\nu} = \partial^{\mu} J^{\nu} - \partial^{\nu} J^{\mu},
  \label{22}
\end{equation}
where $J^{\mu}$ is a 4 - vector (or a 4 - pseudovector). In this
case the amplitude (\ref{20}) is 
\begin{equation}
  A = J^{\mu} (-k) D_{\mu\nu} (k) J^{\nu} (k), 
  \label{23}
\end{equation}
where (see Ref. \cite{1})
\begin{equation}
  D_{\mu\nu} (k) = g_{\mu\nu} - \frac{k_{\mu} k_{\nu}}{k^{2}}.
  \label{24}
\end{equation}
To understand this result we observe that from the field equation 
\[
  \partial_{\mu} G_{\nu} - \partial_{\nu} G_{\mu} = - \partial_{\mu}
  J_{\nu} + \partial_{\nu} J_{\mu}
\]
we get 
\[
  G^{\mu} = - J^{\mu} - \partial^{\mu} \varphi,
\]
where $\varphi (x)$ is a scalar field. The interaction term
$\frac{1}{2} j^{\mu\nu} B_{\mu\nu}$ with the current (\ref{22}) is 
equivalent to 
\[
  - J_{\mu} G^{\mu} = J_{\mu} J^{\mu} + \partial_{\mu} \varphi
   J^{\mu},
\]
what is in agreement with Eqs \ (\ref{23}) and (\ref{24}). So, the 
notoph is propagated if the current $J^{\mu}$ is not conserved. The
process of the notoph radiation (or absorption) is possible too, if 
the current does not conserve. Indeed, such processes are described 
by the vertex $J_{\mu} \partial^{\mu} \varphi$.

Interaction of the notoph with other fields (via the strong conserving
current) have been discussed in Ref. \cite{1}. To illustrate the 
main features of such interactions we consider, for example, the 
interaction with the Dirac field $\psi$ 
\[
  {\cal L} = - \frac{1}{2} G_{\mu} G^{\mu} - J_{\mu} G^{\mu} + 
  \bar{\psi} (- i \gamma_{\mu} \partial^{\mu} + m) \psi
\]
where $J^{\mu} = \bar{\psi} \gamma^{\mu} \psi$ or $J^{\mu} = 
\bar{\psi} \gamma^{\mu} \gamma_{5} \psi$. In the first case the
notoph is not radiated and propagated: the local $J^{\mu} J_{\mu}$ 
interaction appears. In the second case the notoph can be radiated 
and propagated: the local term $J^{\mu} J_{\mu}$ appears too.

\bigskip  We are grateful to Prof. J. Rembieli\'nski and Dr B. Broda
for interesting discussions.

\newpage

\end{document}